\documentclass[aps,pra,reprint]{revtex4-2}

\usepackage{amsmath,amsfonts,amssymb,graphicx,braket,bbold,mathtools,array}
\usepackage{float}
\usepackage{multirow}
\usepackage[]{units}
\usepackage{hyperref}
\usepackage{MnSymbol}
\hypersetup{
    colorlinks=true,
    linkcolor=blue,
    filecolor=blue,      
    urlcolor=blue,
    citecolor=blue
}

\usepackage{braket}
\usepackage[table]{xcolor}

\usepackage[T1]{fontenc}
\usepackage{tikz}
\usetikzlibrary{calc,patterns,decorations.pathmorphing,decorations.markings, shapes, positioning, shapes.geometric}

\newcommand*{\figref}[2][]{%
  \hyperref[{fig:#2}]{%
   \ref*{fig:#2}%
    \ifx\\#1\\%
    \else #1%
    \fi
  }%
}

\begin{document}
\title{Transmission-based noise spectroscopy for quadratic qubit-resonator interactions}
\author{Philipp M. Mutter}
\email{philipp.mutter@uni-konstanz.de}
\author{Guido Burkard}
\email{guido.burkard@uni-konstanz.de}
\affiliation{Department of Physics, University of Konstanz, D-78457 Konstanz, Germany}

\begin{abstract}
We develop a theory describing the transient transmission through noisy qubit-resonator systems with quadratic interactions as are found in superconducting and nanomechanical resonators coupled to solid-state qubits. After generalizing the quantum Langevin equations to arbitrary qubit-resonator couplings, we show that only the cases of linear and quadratic couplings allow for an analytical treatment within standard input-output theory. Focussing for the first time on quadratic couplings and allowing for arbitrary initial qubit coherences, it is shown that noise characteristics can be extracted from input-output measurements by recording both the averaged fluctuations in the transmission probability and the averaged phase. Our results represent an extension to the field of transmission-based noise spectroscopy with immediate practical applications.

\end{abstract}

\maketitle

\section{Introduction}
We live in the age of noisy intermediate-scale quantum devices in which noise severely affects the coherence of qubits and thereby the performance of quantum computers~\cite{Preskill2018}. The noise can have several origins but may roughly be categorized according to whether it arises as a consequence of imperfect system control or unwanted interactions with the environment. For instance, random electromagnetic fields due to fluctuating control gate voltages belong to the former class~\cite{Jirovec2021}, while noise arising from various two-level fluctuators in the host material~\cite{Paladino2014review}, the interaction with the nuclear spin bath in semiconductor quantum dots~\cite{Kloeffel_review2013, Zhang_review2019, Burkard2021arXiv}, as well as magnetic flux noise and quasiparticles in superconducting circuits \cite{Clarke2008,Oliver2013,Krantz2019,deGraaf2020} can be assigned to the latter class.
	
Common to all types of noise is their detrimental effect on the coherence of engineered quantum systems. To mitigate noise-induced decoherence one requires knowledge of the spectral form of the fluctuations which is encoded in the frequency-dependent noise power spectral density $S(\omega)$~\cite{Forster2014, Shulman2014}. Experiments assessing the effect of charge noise on a semiconductor double quantum dot by examining the long-time resonator transmission have already been performed~\cite{Basset2014}, and recently spectroscopy methods for classical and quantum noise based on measurements of the transient transmission through a linearly coupled qubit-resonator system were developed~\cite{Mutter2022, McIntyre2022}. Linear light-matter interactions involving the exchange of one photon are relevant in many fundamentally and technologically relevant systems, such as atoms in a cavity~\cite{Mabuchi2002, Walther2006} as well as charge qubits coupled to superconducting transmission lines~\cite{Wallraff2004, Schuster2005}. Additionally, linear spin-photon couplings are made possible by the spin-orbit interaction~\cite{Kloeffel2013b,Mutter2020cavitycontrol, Mutter2021natural, Mutter2021_ST_qubit, Jirovec2022} or the use of micromagnets~\cite{Benito2017, Mi2018}.

	\begin{figure}
		\includegraphics[scale=0.34]{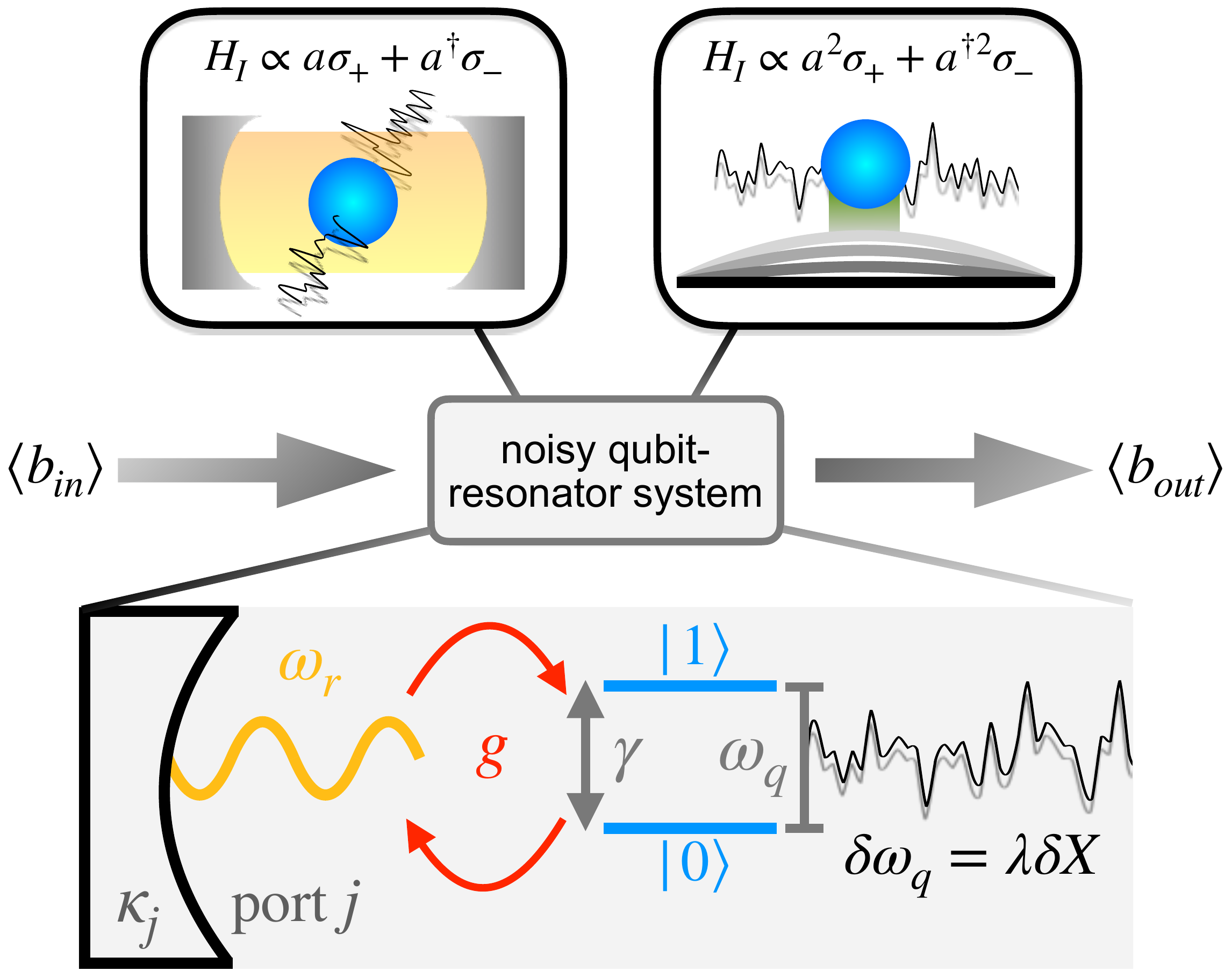}
			\caption{Potential systems for transmission-based noise spectroscopy. A two-level system (qubit, blue) is affected by noise leading to fluctuations $\delta\omega_q$ in the energy splitting $\omega_q$ between qubit states $|0\rangle$ and $|1\rangle$. The interaction with a single resonator mode is described by the Hamiltonian $H_I$ containing the qubit and resonator operators $\sigma_\pm$ and $a$. We schematically show an optical resonator with linear qubit-mode couplings on the left and a nanomechanical resonator with quadratic qubit-mode couplings on the right. Both types of interactions allow for the characterization of noise features by studying the transmission $A = \langle b_{\text{out}} \rangle /\langle b_{\text{in}} \rangle $ through the noisy qubit-resonator system. The bottom schematic introduces the photon loss rate $\kappa_j$ at the resonator port $j$, the qubit decay rate $\gamma$ and the qubit-resonator coupling strength $g$.}
		\label{fig:quad:system_schematic}
	\end{figure}

Quadratic qubit-resonator couplings also allow for input-output measurements and are found in diverse physical systems. For instance, nanomechanical and optomechanical resonators have been shown to couple to superconducting and spin qubits, and the characteristic feature is the quadratic qubit-resonator interaction including two-photon processes~\cite{Zhou2006, Hauss2008, Sankey2010,Wang2016, Munoz2018}. Quadratic qubit-resonator interactions can also be engineered in superconducting circuits~\cite{Kim2015,Felicetti2018b}, and earlier studies of such systems investigated thermal transport~\cite{Wang2021} and input-output theoretical transmission features in the noise-free steady-state case~\cite{Felicetti2018}. If one is interested in transmission-based noise spectroscopy, it is desirable to have a detailed knowledge about the input-output behaviour of both the linear and quadratic cases as the coupling of a given type of qubit to resonator modes may more easily be realized in one system or the other. In this paper we generalize the results in Ref.~\cite{Mutter2022} to also include quadratic qubit-resonator interactions and show that noise characteristics can be inferred by measuring both the averaged fluctuations in the transmission probability $\delta \llangle \vert A \vert^2 \rrangle/ \vert A_{\infty} \vert^2$ and the averaged phase $\llangle \phi \rrangle$. Moreover, as another extension of earlier results we allow for a general form of the initial qubit coherence $\langle \sigma_-^0 \rangle$.

The remainder of this paper is structured as follows: In Sec.~\ref{sec:quad_model_and_Langevin_equations} we consider a general $n$-photon-qubit interaction and derive the Langevin equations for this case. We then show that within standard input-output theory only the cases $n=1$ and $n=2$ allow for an analytical solution to lowest order in perturbation theory. Specializing on these cases we calculate the averaged transmission probability and phase in Sec.~\ref{sec:quad_transmission}, and propose a scheme for extracting noise characteristics in Sec.~\ref{sec:quad:extracting_noise_characteristics}. Finally, Sec.~\ref{sec:conclusion} provides a conclusion.

\section{General model and Langevin equations}
\label{sec:quad_model_and_Langevin_equations}
We consider a resonator mode of frequency $\omega_r$ that couples to a qubit with fluctuating energy separation $\omega_q + \delta \omega_q(t)$, see Fig.~\ref{fig:quad:system_schematic} (bottom). The noise $\delta X$ affecting the control parameter $X$ is related to the fluctuating energy separation by the first-order expansion $\delta \omega_q = \lambda \delta X$ with the noise sensitivity $\lambda = \partial_X \omega_q \vert_{\delta X = 0}$. The number $n$ of photons taking part in the interaction is left unspecified for now, and we assume a general qubit-resonator Hamiltonian of the form 
	\begin{align}
		H_{qr} = \frac{\omega_q + \delta \omega_q(t)}{2} \sigma_z + \omega_r a^{\dagger} a + g(a+a^{\dagger})^n \sigma_x,
	\end{align}
where $\sigma_{x,z}$ are Pauli matrices in the basis of the logical qubit states $\lbrace \vert 0 \rangle, \vert 1 \rangle \rbrace$, and $a$ is the mode annihilation operator. In the rotating frame defined by the transformation $H_{qr} \rightarrow U H_{qr} U^{\dagger} + i \dot{U} U^{\dagger}$ with the time-dependent unitary $U = \exp (i \Omega t [a^{\dagger}a/n + \sigma_z /2])$ and within the rotating wave approximation, the system is described by the Hamiltonian
	\begin{align}
	\label{eq:H_RWA}
		\mathcal{H}_{qr} = \frac{\Delta_q + \delta \omega_q(t)}{2} \sigma_z + \Delta_r^{(n)} a^{\dagger} a + g \left[ a^n \sigma_+ + a^{\dagger n} \sigma_- \right],
	\end{align}
where  $\sigma_{\pm}$ are the qubit ladder operators, and the quantities $\Delta_q = \omega_q -  \Omega$ and $\Delta_r^{(n)} = \omega_r - \Omega/n$ are the shifted qubit and resonator frequencies.

In addition, external bath modes $b$ interact with the resonator, and the coupling is assumed to be linear. We take into account the effects of the corresponding input field $\langle b_{\text{in}}(t) \rangle$ by including the injection Hamiltonian in the rotating frame,
    \begin{align}
    \label{eq:H_in}
        \mathcal{H}_{\text{in}} = i \sqrt{\kappa_1}  \left( \langle b_{\text{in}}(t) \rangle e^{i \Omega t/n} a^{\dagger} - \text{H.c.} \right),
    \end{align}
The input field $\langle b_{\text{in}} (t) \rangle$ may be used to probe the system, and it is assumed to be a classical  wave of real amplitude $\langle b_{\text{in}}  \rangle$ and frequency $\omega_p$, $\langle b_{\text{in}}  (t)\rangle = \langle b_{\text{in}}  \rangle e^{- i \omega_p t}$. In the following we work in the frame defined by $U$ as above with $\Omega = n \omega_p$ such that the the injection Hamiltonian~\eqref{eq:H_in} is constant in time. One then has $\Delta_q \equiv \Delta_q^{(n)} = \omega_q - n \omega_p$ and $\Delta_r^{(n)} \equiv \Delta_r = \omega_r - \omega_p$.

Furthermore taking into account noise-independent qubit relaxation, absorption and dephasing at finite temperature as well as photon losses, the system may be described by a Lindblad master equation in the rotating frame,
    \begin{align}
    \label{eq:Lindblad}
    \begin{split}
         \dot{ \rho}&=  -i \left[\mathcal{H}_{qr} + \mathcal{H}_{\text{in}}, \rho \right] + \mathcal{L}[\rho], \\
          \mathcal{L}[\rho] &=  \frac{\gamma_1}{2} \sum_{\pm} \frac{1 \pm 1 + 2n_B}{2} \left(2 \sigma_{\mp} \rho \sigma_{\pm} -  \lbrace \rho,  \sigma_{\pm} \sigma_{\mp} \rbrace \right) \\          
         & \quad  + \frac{\gamma_{\varphi}}{2} \left( \sigma_z \rho \sigma_z - \rho \right) 
          + \frac{\kappa}{2} \left( 2 a \rho a^{\dagger} - \lbrace \rho,  a^{\dagger} a \rbrace \right),
    \end{split}
    \end{align}
where $\gamma_1$ is the relaxation rate at zero Kelvin, $\gamma_{\varphi}$ is the dephasing rate and $n_B = [\exp( \omega_q/T) -1]^{-1}$ is the Bose distribution, i.e., the mean number of bath photons at temperature $T$, evaluated at the qubit frequency where resonant energy exchange is possible. Additionally, we have introduced the total photon loss rate $\kappa = \sum_j \kappa_j + \kappa_{\text{int}}$ as the sum of the individual loss rates at port $j \in \lbrace 1,2 \rbrace$ and the internal loss rate $\kappa_{\text{int}}$.

By using the relation $\langle \dot{O} \rangle = \text{Tr} (\dot{\rho} O)$ for a given operator $O$ and Eq.~\eqref{eq:Lindblad}, one obtains the partial system of Langevin equations to leading order in $g$,
	\begin{align}
	\label{eq:quad_Langevin_equations}
	\begin{split}
	\dot{\langle a \rangle} & = - \left[i \Delta_r + \frac{\kappa}{2} \right] \langle a \rangle - i g n \langle \sigma_- \rangle \langle a^{n-1} \rangle^* + \sqrt{\kappa_1} \langle b_{\text{in}} \rangle, \\
		\dot{\langle \sigma_z \rangle}  & =  - \gamma_1 (T) \langle \sigma_z \rangle -  \gamma_1(0) + 2ig  \left(\langle a^n \rangle \langle \sigma_- \rangle^*  -  \text{c.c.}  \right)    ,  \\
		\dot{ \langle \sigma_-  \rangle }  & = - \left[ i\Delta_q^{(n)} + i \lambda \delta X(t) + \gamma_2\right]  \langle \sigma_- \rangle  + ig \langle \sigma_z \rangle \langle a^n \rangle ,
	\end{split}
	\end{align}
where $\gamma_1(T) = \gamma_1 \coth(\omega_q /2T)$,  $\gamma_2 =  \gamma_1(T)/2 +  \gamma_{\varphi}$, the star denotes complex conjugation, and we have used the relation $\langle Q R \rangle = \langle Q \rangle \langle R \rangle + \mathcal{O}(g)$ that is valid for all qubit (Q) and resonator (R) operators under the assumption of an initially separable state. In the following we define $\gamma = 2 \gamma_2$ such that the homogeneous parts of the Langevin equations for $\langle \sigma_- \rangle$ and $\langle a \rangle$ have the same form, allowing us to obtain cleaner mathematical expressions later on.

\subsection{Analytical approach}
\label{subsec:analytical_approach}

The system of Langevin equations~\eqref{eq:quad_Langevin_equations} is not closed because in general $\langle a^n \rangle \neq \langle a \rangle^n$, and even if it the relation would hold true, an exact solution in the presence of arbitrary time-dependent qubit noise could not be obtained. However, we may formally solve the equation for $\langle \sigma_- \rangle$ without any assumptions on the noise $\delta X(t)$,
	\begin{align}
		&\langle \sigma_-(t) \rangle =   \langle \sigma_-^0 \rangle e^{-i \Delta_q^{(n)} t - \gamma t/2} e^{-i \lambda \mathcal{X}(t)}   \\
		 &+ i g \int_0^t \langle \sigma_z (t') \rangle \langle a^n(t') \rangle e^{(i \Delta_q^{(n)} + \gamma /2)(t'-t)} e^{i \lambda [\mathcal{X}(t')-  \mathcal{X}(t)]} dt' \nonumber ,
	\end{align} 
with the integrated noise $\mathcal{X}(t) = \int_0^t  \delta X(t') dt'$ and the initial qubit coherence $\langle \sigma_-^0 \rangle$, and subsequently insert the result into the equation for $\langle a \rangle$. By again formally solving the resulting differential equation, we find an integral equation for the expectation value of the mode annihilation operator,
	\begin{widetext}
	\begin{align}
	\label{eq:quad:a_general}
		\langle a(t) \rangle      & =  \frac{ \sqrt{\kappa_1}  \langle b_{\text{in}} \rangle}{i \Delta_r +  \kappa/2} \left(1 - e^{-i \Delta_r t -\kappa t /2} \right)   -  i n g \langle \sigma_-^0 \rangle e^{-i \Delta_r t -\kappa t /2} \int_0^t d t^{\prime}  \langle a^{n-1} (t') \rangle^* e^{i (\Delta_r - \Delta_q^{(n)}) t^{\prime} + (\kappa - \gamma) t^{\prime}/2} e^{-i \lambda \mathcal{X}(t^{\prime})}    \\
		&  + n g^2 e^{-i \Delta_r t -\kappa t /2} \int_0^t d t^{\prime}  \langle a^{n-1} (t') \rangle^* e^{i (\Delta_r - \Delta_q^{(n)})  t^{\prime} + (\kappa - \gamma) t^{\prime}/2} e^{-i  \lambda \mathcal{X}(t^{\prime})}   \int_0^{t^{\prime}}   d t^{\prime \prime} \langle \sigma_z (t^{\prime \prime}) \rangle \langle a^n(t^{\prime \prime}) \rangle  e^{i \Delta_q^{(n)}  t^{\prime \prime} + \gamma t^{\prime \prime}/2} e^{-i \lambda \mathcal{X}(t^{\prime \prime})}.  \nonumber
	\end{align}
	\end{widetext}
where we assume $ \langle a^0 \rangle =0$, e.g., the resonator may initially be in an eigenstate of the mode number operator $a^{\dagger} a$. We first remark that if we wish to work to leading order in $g$, we may neglect the second integral in~\eqref{eq:quad:a_general}, and hence the time-dependent populations $\langle \sigma_z (t) \rangle$ do not play a role. Secondly, if $n \leqslant 2$, then the $\langle a^{n-1} \rangle$ term in the remaining integral is equal to unity  ($n=1$) or $\langle a \rangle$ ($n=2$), and the equation can be solved to first order in perturbation theory by substituting the zeroth-order value of $\langle a \rangle$ into the first integral. Otherwise, the fact that in general $\langle a^{n-1} \rangle \neq \langle a \rangle^{n-1} $ forces us to obtain an additional equation of motion for $\langle a^{n-1} \rangle$ which cannot be treated in a simple input-output model without further assumptions due to additional terms of the form $\int  \langle b(\omega) a^{n-2} \rangle d \omega$ which feature the bath modes $b(\omega)$ but cannot be related to the input fields. In the long-time limit the term linear in $g$ vanishes, and the long-time transmission is determined by the $g^2$ term. Since this term features the quantity $\langle a^n \rangle$, the long-time solution can only be obtained for $n=1$ due to the same arguments made above. A comprehensive study of the long-time transmission for linearly coupled systems may be found in Ref.~\cite{Mutter2022b}. Since the coupling of the resonator to the external bath modes is assumed to be linear, the standard input-output relation for the transmission amplitude $A$ applies~\cite{Collett1984, Gardiner1985, Burkard2020}, 
	\begin{align}
	\label{eq:transmission_amplitude_general}
		A(t) = \frac{\langle b_{\text{out}} (t) \rangle  }{ \langle b_{\text{in}} (t)\rangle} =  - \frac{\sqrt{\kappa_2} \langle a (t) \rangle  }{ \langle b_{\text{in}}(t) \rangle}.
	\end{align}
As a consequence, the transmission amplitude can be obtained by solving Eq.~\eqref{eq:quad:a_general} for $\langle a (t) \rangle$.

\subsection{Numerical study}
\label{subsec:numerical_study}

To validate the truncation of the quantum Langevin equations~\eqref{eq:quad_Langevin_equations} and the solution for the transmission at first order in the qubit-photon coupling $g$, we numerically solve the full Lindblad equation~\eqref{eq:Lindblad} and compare the results to our approximate analytical expressions in Figs.~\ref{fig:averaged_transmission} and~\ref{fig:averaged_phase}. We allow for up to $12$ photons and assume an initially empty resonator containing a qubit in a coherent superposition $\vert \psi \rangle = (e^{i\pi/8} \ket{ \uparrow } + e^{-i\pi/8}\ket{ \downarrow })/\sqrt{2}$ such that the initial conditions $\langle a^0 \rangle = 0$, $\text{Re} \langle \sigma_-^0 \rangle = \text{Im} \langle \sigma_-^0 \rangle = 1/\sqrt{8} $ and $\langle \sigma_z^0 \rangle = 0$ are satisfied. The transmission amplitude is then calculated according to Eq.~\eqref{eq:transmission_amplitude_general} as $A = - \sqrt{\kappa_2} \text{Tr}(\rho a)/ \langle b_{\text{in}} \rangle $, and the averaged transmission and phase are obtained by averaging the solution over $10^3$ distinct noise configurations. We generally find excellent agreement in the regime considered, underlining the high accuracy of our analytical results. 

\section{Observable figures of merit} \label{sec:quad_transmission}
We now focus on the cases $n \leqslant 2$ that can be treated perturbatively. The case $n=1$ has been studied in Ref.~\cite{Mutter2022}, and we include it here to be able to make contact to these previous considerations. In addition, we complement earlier results by including the averaged phase $\llangle \phi \rrangle$ as an observable figure of merit and allowing for arbitrary complex initial qubit coherences $\langle \sigma_-^0 \rangle$. The case $n=2$ constitutes a completely new contribution to the theory of transmission-based noise spectroscopy, and it is of immediate interest for nanomechanical and superconducting resonators coupled to solid-state qubits. 

The complex transmission amplitude can be written as 
	\begin{align}
	\label{eq:transmission_amplitude_complex}
		A = \vert A \vert e^{i \phi},
	\end{align}
where both the transmission $\vert A \vert$ and the phase $\phi$ are experimentally observable and thus allow for a noise average $\llangle \dots \rrangle$ over many measurements. Solving Eq.~\eqref{eq:quad:a_general} to first order in the small parameter $\varepsilon^{(n)} \equiv n g  / \text{max} \lbrace \vert \delta_0^{(n)} \vert \equiv \vert \Delta_r - \Delta_q^{(n)} \vert, \vert \kappa - \gamma \vert \rbrace$, we find according to Eq.~\eqref{eq:transmission_amplitude_general},
	\begin{align}
	\label{eq:transmission_amplitude_n}
		\frac{A_n(t)}{A_{\infty}}  & =    1 - e^{-i \Delta_r t- \kappa t/2} \\
		& \times \bigg[1 + ign \langle\sigma_-^0 \rangle    \frac{\kappa/2 + i \Delta_r}{\kappa/2 - i \Delta_r} \left(  \frac{ \kappa/2 - i \Delta_r }{ \sqrt{\kappa_1} \langle b_{\text{in}} \rangle} \right)^{2-n} \mathcal{I}_n(t) \bigg] \nonumber,
	\end{align}
with the zeroth-order long-time transmission amplitude $A_{\infty} = - \sqrt{\kappa_1 \kappa_2}/(i \Delta_r + \kappa/2)$ and the noise integrals
	\begin{align}
	\begin{split}
	\label{eq:total_noise_integral_n}
		\mathcal{I}_n(t)  = & \int_0^t e^{i(\omega_r - \omega_q) t'  + (\kappa- \gamma) t'/2} \\
		&\qquad  \times \left( e^{i  \omega_p t' } - e^{i \omega_r   t' - \kappa t'/2} \right)^{n-1} e^{-i \lambda \mathcal{X}(t') } dt'.
	\end{split}
	\end{align}
Interestingly, the corrections to the empty resonator transmission come with a factor $ \langle b_{\text{in}} \rangle^{n-2}$, and hence the transient transmission for linear interactions depends on the input field while for quadratic interactions it does not, a characteristic feature of this type of qubit-resonator coupling. To extract noise characteristics, we now proceed to study the averaged transient transmission and phase.

\subsection{Transmission}
\label{subsec:transmission}

	\begin{figure}
		\includegraphics[scale=0.255]{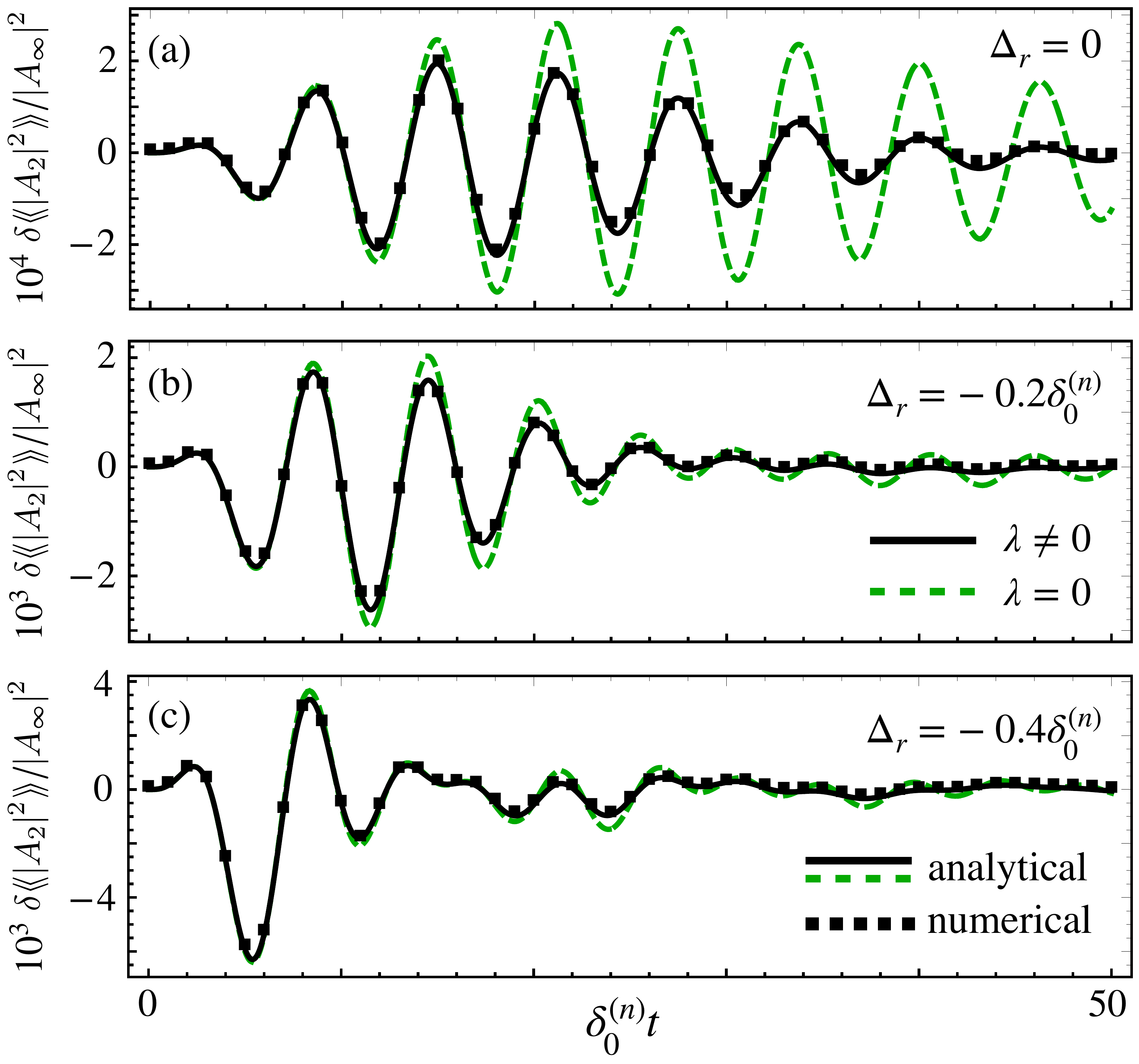}
			\caption{The normalized fluctuations in the averaged transmission probability $\delta \llangle \vert A_n \vert^2 \rrangle /\vert A_{\infty} \vert^2 $ for quadratic qubit-resonator couplings ($n=2$) as a function of time $t$. Solid lines are drawn according to Eq.~\eqref{eq:individual_terms} with the ANI for quasistatic noise [Eq.~\eqref{eq:ANI_2_qs_noise}] with root-mean-square $\delta X_{\text{rms}} = 0.05 \delta_0^{(n)}$, while the numerical data points (squares) are obtained as described in Sec.~\eqref{subsec:numerical_study}. The parameter values set to $n g =  0.002 \delta_0^{(n)}$, $\kappa =  0.1 \delta_0^{(n)}$, $\kappa_1 = \kappa_2 = \kappa/2$, $\kappa_{\text{int}} = 0$, $\gamma_1 = \gamma_{\varphi} = 0.025\delta_0^{(n)}$, $T=\delta_0^{(n)}$, $\lambda = 0.9$, and the resonator-probe detunings are chosen as indicated in the figure.}
			\label{fig:averaged_transmission}
	\end{figure}
	
Taking the absolute value of~\eqref{eq:transmission_amplitude_n} and averaging over distinct noise configurations yields
	\begin{align}
	\label{eq:averaged_transmission_n}
		\llangle \vert A_n \vert \rrangle = \vert A_{\infty} \vert \sqrt{\xi_0 + \xi_{1,n}},
	\end{align}
where $\vert A_{\infty} \vert = \sqrt{\kappa_1 \kappa_2/(\Delta_r^2 + \kappa^2/4)}$ is the long-time empty cavity transmission,
	\begin{align}
	\label{eq:individual_terms}
		 \xi_0(t) & =    1 + e^{- \kappa t} - 2 e^{-\kappa t/2} \cos (\Delta_r t) ,  \nonumber  \\
		 \xi_{1,n} (t) & = 2ng  e^{-\kappa t/2} \left( \frac{\sqrt{\Delta_r^2 + \kappa^2/4}  }{\sqrt{\kappa_1} \langle b_{\text{in}} \rangle} \right)^{2-n}  \\
		 & \times   \bigg( Y_n(t) \left[ \text{Re} \langle \sigma_-^0 \rangle \text{Re} \llangle \mathcal{I}_n \rrangle - \text{Im} \langle \sigma_-^0 \rangle \text{Im} \llangle  \mathcal{I}_n \rrangle \right] \nonumber \\
		 &\quad  +Z_n(t) \left[  \text{Im} \langle \sigma_-^0 \rangle \text{Re} \llangle  \mathcal{I}_n \rrangle + \text{Re} \langle \sigma_-^0 \rangle \text{Im} \llangle \mathcal{I}_n \rrangle  \right]   \bigg) \nonumber,
	\end{align}
and we have introduced the functions
	\begin{align}
	\begin{split}
		Y_n(t) & = \sin (n \zeta - \Delta_r t) - e^{-\kappa t/2} \sin (n \zeta), \\
		Z_n(t) & = \cos (n \zeta - \Delta_r t) - e^{-\kappa t/2} \cos (n \zeta),
	\end{split}
	\end{align}
with $\zeta = \arctan (2 \Delta_r /\kappa)$. In obtaining~\eqref{eq:averaged_transmission_n} we have used the fact that the variance of $\vert A \vert$ is non-zero only at quadratic order in $\varepsilon^{(n)}$ or higher~\cite{Mutter2022}. The averaged noise integrals (ANIs) $\llangle \mathcal{I}_n \rrangle$ are as given in Eq.~\eqref{eq:total_noise_integral_n} but with the factor $\exp (-i \lambda \mathcal{X}(t') )$ in the integral replaced by the averaged random phase $\llangle \exp (-i \lambda \mathcal{X}(t') ) \rrangle$. One may rewrite Eq.~\eqref{eq:averaged_transmission_n} in terms of the normalized fluctuations of the averaged transmission probability,
	\begin{align}
	\label{eq:quad_averaged_subtracted}
		\frac{\delta 	\llangle \vert A_n \vert^2 \rrangle}{\vert A_{\infty} \vert^2} = \frac{ 	\llangle \vert A_n \vert^2 \rrangle - \xi_0 \vert A_{\infty} \vert^2}{\vert A_{\infty} \vert^2} = \xi_{1,n}.
	\end{align}
The effect of the noise on the averaged transmission can be assessed in a clear and disentangled fashion by using Eq.~\eqref{eq:quad_averaged_subtracted} as can be seen from Fig.~\ref{fig:averaged_transmission}.

Often the resonator is probed on resonance, $\Delta_r = 0$, since one may expect the most pronounced response at this point in experiments. In this case, one finds
	\begin{align}
	\label{eq:transmission_probability_resonance}
	\begin{split}
		\frac{\delta 	\llangle \vert A_n \vert^2 \rrangle}{\vert A_{\infty} \vert^2}  =   2&ng \left(\frac{\kappa  }{2 \sqrt{\kappa_1} \langle b_{\text{in}} \rangle} \right)^{2-n} e^{-\kappa t/2}   \left( 1 - e^{-\kappa t/2} \right)\\
		& \times \left[ \text{Im} \langle \sigma_-^0 \rangle \text{Re} \llangle  \mathcal{I}_n \rrangle  + \text{Re} \langle \sigma_-^0 \rangle \text{Im} \llangle \mathcal{I}_n \rrangle \right],
	\end{split}
	\end{align}
where $ \vert A_{\infty} \vert =1$ for symmetric mirrors. The transient signal for the resonant case is shown in Fig.~\figref[(a)]{averaged_transmission}.

\subsection{Phase}	
\label{subsec:phase}
	
\begin{figure}
		\includegraphics[scale=0.275]{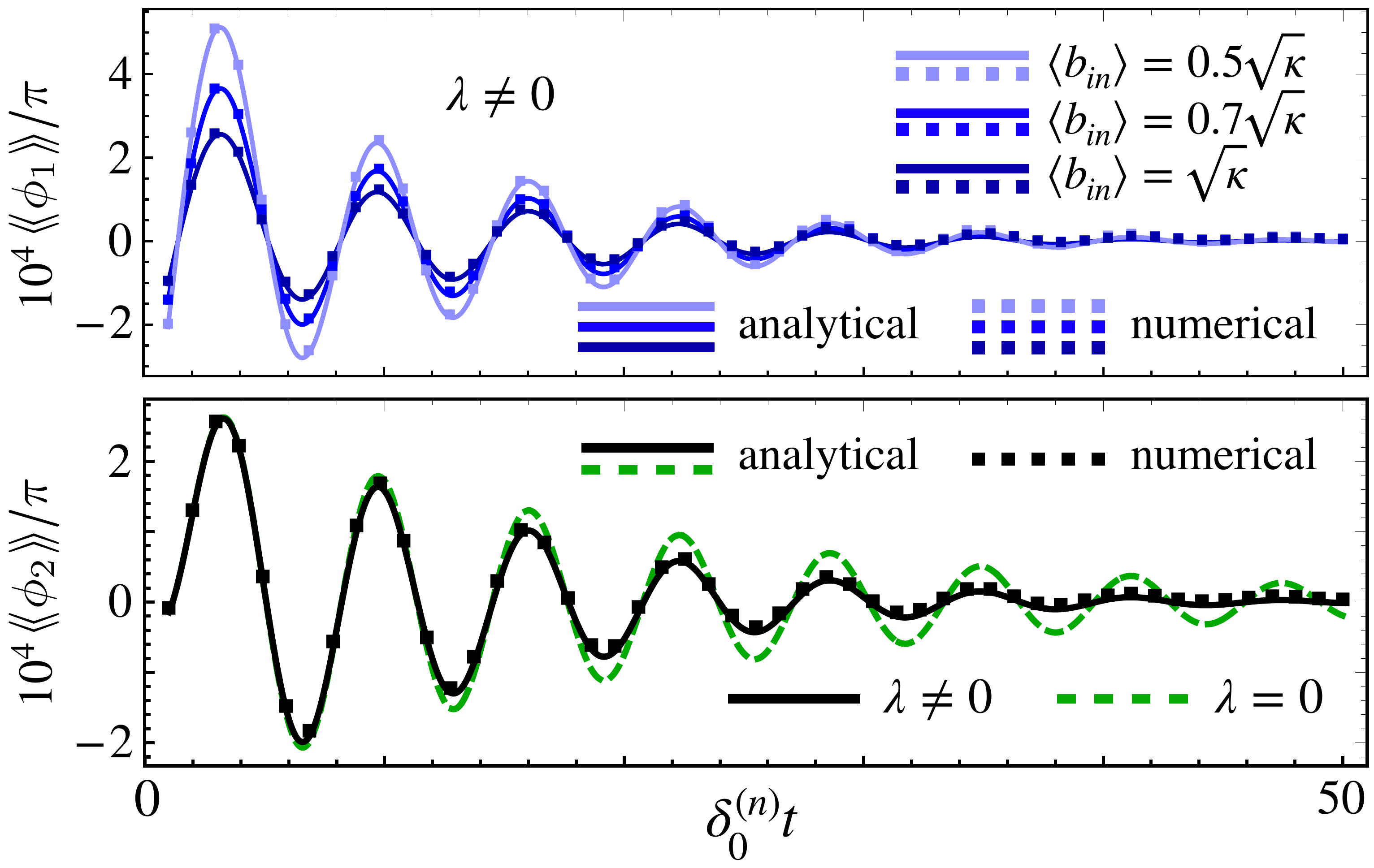}
			\caption{The transient averaged phase $\llangle \phi_n \rrangle$ as a function of the measurement time $t$ for quasistatic noise at $\Delta_r = 0$ for the cases of linear qubit-resonator couplings ($n=1$, top) and quadratic qubit-resonator couplings ($n=2$, bottom). The solid lines are drawn according to the approximate analytical result~\eqref{eq:phase_expressions}, while the numerical data is obtained as described in Sec.~\ref{subsec:numerical_study} using the exact expression~\eqref{eq:averaged_phase}. For the case $n=1$ we show the averaged phase for several values of the input field $\langle b_{\text{in}} \rangle$ as indicated in the figure. We set $\kappa = \delta_0^{(n)}$ and the remaining parameter values are chosen as in Fig.~\figref[(a)]{averaged_transmission}. }
		\label{fig:averaged_phase}
	\end{figure}

We now turn to the averaged phase of the transmission signal, which has not been investigated in the context of noise spectroscopy so far. One has
	\begin{align}
	\label{eq:averaged_phase}
		\llangle \phi_n \rrangle = \left\llangle \arctan \left( \frac{\text{Im} (A_n) }{\text{Re} (A_n)} \right) \right\rrangle,
	\end{align}
where the general expressions for the real and imaginary parts of the transmission amplitude to leading order in $g$ are straightforward to calculate from Eq.~\eqref{eq:transmission_amplitude_n} but lengthy, and we display them in Appendix~\ref{appx:real_and_imaginary_parts_of_A}. In the above form the noise average of the phase cannot be performed analytically due to the $\arctan$ function. However, when the resonator is driven on resonance, $\Delta_r=0$, the dominant term in the expansion of $ \text{Im} (A_n) $ in orders of $\varepsilon^{(n)}$ vanishes, while it is non-zero in the expansion of $ \text{Re} (A_n) $, and one has $ \vert \text{Im} (A_n) \vert \ll \vert \text{Re} (A_n) \vert$ for $g \ll \kappa$. In this case, one may keep only the leading-order term in the Taylor expansion of the $\arctan$ function, whereby it becomes possible to average the phase analytically,
	\begin{align}
	\label{eq:phase_expressions}
	\begin{split}
		\llangle \phi_n \rrangle \approx    & \frac{n g }{1 - e^{\kappa t/2}}  \left(\frac{\kappa  }{2 \sqrt{\kappa_1} \langle b_{\text{in}} \rangle} \right)^{2-n} \\
		& \times  \left[ \text{Re} \langle \sigma_-^0 \rangle \text{Re} \llangle \mathcal{I}_n \rrangle - \text{Im} \langle \sigma_-^0 \rangle \text{Im} \llangle \mathcal{I}_n \rrangle \right].
	\end{split}
	\end{align}
As was the case for the corrections to the empty resonator transmission probability, the phase is also inversely proportional to the input field $\langle b_{\text{in}} \rangle$ for linear couplings but independent of the input field for quadratic couplings. To underline the validity of the approximation made when deriving~\eqref{eq:phase_expressions}, we show a comparison between numerical and analytical results for the transient averaged phase in Fig.~\ref{fig:averaged_phase}. We find excellent agreement, suggesting that the averaged phase measured at the resonance $\Delta_r = 0$ can be used to infer noise characteristics according to Eq.~\eqref{eq:phase_expressions}.

\section{Noise spectroscopy}
\label{sec:quad:extracting_noise_characteristics}

The aim of noise spectroscopy is the determination of spectral features of the fluctuations affecting the system under consideration~\cite{Degen2017}. In this section we show that information on the noise power spectral density $S(\omega)$ of the qubit noise can be obtained from the averaged transmission and phase by extracting the ANI and investigating its properties.

\subsection{Extracting the averaged noise integral}
\label{subsec:extracting the noise integral}
Having obtained expressions for the averaged transmission and averaged phase, we now turn to investigate how noise features can be extracted from the measurement of these quantities. For $n=1$, the extraction of the ANI and the extraction of noise features from the ANI is detailed in Ref.~\cite{Mutter2022}. The former relies on the fact that the ANI does not depend on the probe frequency $\omega_p$, allowing for an extraction scheme based on measurements of the transmission at two distinct resonator-probe detunings. For $n=2$, however, the noise integral $\mathcal{I}_2$ does depend on $\omega_p$, and hence it is impossible to extract it by recording the transmission at two distinct values of the probe frequency. However, if in addition the phase is measured, it is possible to obtain the ANI for both $n=1$ and $n=2$. Since we aim to use the analytical result for the averaged phase in Eq.~\eqref{eq:phase_expressions}, we work in the resonant regime where $\Delta_r $ is fixed to zero. One may then infer the following set of equations from Eqs.~\eqref{eq:averaged_transmission_n} and~\eqref{eq:phase_expressions},
	\begin{align}
	\label{eq:system_of_equations}
	\begin{split}
		    \text{Im} \langle \sigma_-^0 \rangle \text{Re} \llangle  \mathcal{I}_n \rrangle + \text{Re} \langle \sigma_-^0 \rangle \text{Im} \llangle \mathcal{I}_n \rrangle   & = \Upsilon_{\vert A \vert }, \\	
		  \text{Re} \langle \sigma_-^0 \rangle \text{Re} \llangle \mathcal{I}_n \rrangle - \text{Im} \langle \sigma_-^0 \rangle \text{Im} \llangle \mathcal{I}_n \rrangle & = \Upsilon_{\phi},
	\end{split}
	\end{align}
where the functions on the right-hand side are obtained from the averaged transmission and phase,
	\begin{align}
		 \Upsilon_{\vert A \vert } & = \frac{\llangle \vert A_n \vert \rrangle^2/\vert A_{\infty} \vert^2 - \left(1- e^{-\kappa t/2} \right)^2}{ 2ng e^{-\kappa t/2} (1 - e^{-\kappa t/2} ) } \left(\frac{\kappa  }{2 \sqrt{\kappa_1} \langle b_{\text{in}} \rangle} \right)^{n-2}, \nonumber  \\
		 \Upsilon_{\phi} & = \llangle \phi_n \rrangle  \frac{1 - e^{\kappa t/2} }{ng}  \left(\frac{\kappa  }{2 \sqrt{\kappa_1} \langle b_{\text{in}} \rangle} \right)^{n-2}.
	\end{align}
The inhomogeneous linear system of equations~\eqref{eq:system_of_equations} can be uniquely solved for the real and imaginary parts of $\llangle \mathcal{I}_n \rrangle$ if
	\begin{align}
		\det \begin{pmatrix}
				\text{Im} \langle \sigma_-^0 \rangle &  	\text{Re} \langle \sigma_-^0 \rangle \\
					\text{Re} \langle \sigma_-^0 \rangle & 	-\text{Im} \langle \sigma_-^0 \rangle
		\end{pmatrix}
		 =   -\left\vert \langle \sigma_-^0 \rangle \right\vert^2 \neq 0.
	\end{align}
Hence, the qubit must be initialized in a coherent superposition to permit the extraction of the ANI from transmission measurements. This condition, however, is not restrictive as it is required to be fulfilled for the noisy part of the transmission signal to be non-vanishing in the transient phase in the first place [see Eq.~\eqref{eq:transmission_amplitude_n}] and hence does not introduce additional constraints.

\subsection{Common types of noise}
\label{subsec:common_types_of_noise}
Since we have shown that the ANI can be extracted from transmission measurements even for quadratic qubit-resonator interactions, it is possible to extract noise features by comparing the measured data to our theory. For this purpose, we now aim to obtain analytical expressions for the ANI for the case $n=2$. To proceed, we assume zero-mean noise and express the factor $\llangle \exp (-i \lambda \mathcal{X}(t') ) \rrangle$ that appears in the ANI in the Gaussian approximation using the noise power spectral density $S(\omega)$~\cite{Bergli2009}. We may then evaluate the ANI for three noise types that are common in engineered quantum systems: white noise, quasistatic noise and low-frequency noise.

White noise is characterized by a constant noise amplitude, $S(\omega ) =S_0$, and the corresponding ANI takes the form
	\begin{align}
	\label{eq:ANI_2_white_noise}
		\llangle \mathcal{I}_2 \rrangle = \sum_{j=1}^2 (-1)^j \frac{e^{c_jt - \lambda^2 S_0 t/2} -1}{c_j - \lambda^2 S_0 /2},
	\end{align}
where we introduce the shorthand notation
	\begin{align}
		c_1 = i(2 \omega_r - \omega_q )  - \frac{\gamma}{2}, \; \; \; c_2 = i(\omega_r - \omega_q + \omega_p) + \frac{\kappa - \gamma}{2}.
	\end{align}		
Fluctuations are quasistatic if they are constant during a given measurement but change between two measurements. In this case, the spectral density may be written as $S(\omega) = 2 \pi \delta X_{\text{rms}}^2 \delta (\omega)$, where $\delta X_{\text{rms}} = \sqrt{\llangle \delta X^2 \rrangle}$ is the root-mean-square of the noise and $\delta(\omega)$ is the Dirac delta distribution, and one finds 
	\begin{align}
	\label{eq:ANI_2_qs_noise}
		& \llangle \mathcal{I}_2 \rrangle = \sum_{j=1}^2  \frac{(-1)^j  e^{C_j^2}}{\lambda \delta X_{\text{rms}} } \left[ \mathcal{E}\left( C_j \right) +  \mathcal{E}\left( \frac{ \lambda \delta X_{\text{rms}} }{\sqrt{2}} t  - C_j \right) \right],
	\end{align}
where $C_j = c_j/(\sqrt{2} \lambda \delta X_{\text{rms}} )$ and $\mathcal{E}(z) = \sqrt{\pi/2}~\text{erf}(z)$ with error function $\text{erf}(z)$. Finally, one may consider low-frequency noise such as $1/f$ noise in a frequency band $[\omega_{\text{ir}}, \omega_{\text{uv}}]$. If large frequencies are sufficiently strongly suppressed such that the ultraviolet cutoff frequency $\omega_{\text{uv}}$ satisfies $\omega_{\text{uv}} t \ll 1 $ in the transient phase, the ANI is in good approximation described by the same functional form as in Eq.~\eqref{eq:ANI_2_qs_noise} with the substitution $\delta X_{\text{rms}} \rightarrow \sqrt{P/\pi}$, where $P = \int S(\omega) d \omega$ is the noise power in the frequency band under consideration. The information on the noise attainable by fitting the experimental data to the above expressions hence includes the noise amplitude $S_0$ for white noise, the root mean square of the noise $ \delta X_{\text{rms}}$ for quasistatic noise and the noise power $P$ for low-frequency noise. While useful when investigating common material platforms, the approach requires a priori knowledge on the type of noise. We now proceed to introduce a formalism that allows the extraction of noise features in terms of the power spectral density $S(\omega)$ without the need to make any assumptions on the form of the noise.

\subsection{Relating the noise integrals}
\label{subsec:relating_the_noise_integrals}
In a final step, we wish to relate the $n=1$ and $n=2$ ANIs since we already know how to extract $S$ from $\llangle \mathcal{I}_1 \rrangle$ via the convolution theorem from the discussion in Ref.~\cite{Mutter2022}. Comparing the expressions for $ \mathcal{I}_1 $ and $ \mathcal{I}_2$ in Eq.~\eqref{eq:total_noise_integral_n}, one straightforwardly finds the relation
	\begin{align}
	\label{eq:relation_ANIs}
			\mathcal{I}_2(t) = \int_0^t \left(e^{i \omega_p t'} - e^{i \omega_r t' - \kappa t'/2} \right)  \dot{\mathcal{I}}_1(t') dt',
	\end{align}
where $\dot{\mathcal{I}}_1(t')  = \partial_t \mathcal{I}_1(t) \vert_{t = t'}$. Eq.~\eqref{eq:relation_ANIs} can be inverted, and after averaging over the noise configurations, we find
	\begin{align}
	\label{eq:relation_ANIs_2}
		& \llangle \mathcal{I}_1(t) \rrangle   =  \int_0^t \frac{ \llangle \dot{\mathcal{I}}_2 (t') \rrangle}{e^{i \omega_p t'} - e^{i \omega_r t' - \kappa t'/2}} dt'  = \frac{\llangle \mathcal{I}_2 (t) \rrangle}{e^{i \omega_p t} - e^{i \omega_r t - \kappa t/2} } \nonumber  \\
		&  + \int_0^t \llangle \mathcal{I}_2 (t') \rrangle \frac{   i \omega_p e^{i \omega_p t'} - \left[i \omega_r - \frac{\kappa}{2} \right] e^{i \omega_r t' - \kappa t'/2}  }{ \left[e^{i \omega_p t'} - e^{i \omega_r t' - \kappa t'/2} \right]^2}   dt' ,
	\end{align}
where the second equality is obtained by partial integration. After converting the ANI $	\llangle \mathcal{I}_2(t) \rrangle$ to $\llangle \mathcal{I}_1(t) \rrangle$ via Eq.~\eqref{eq:relation_ANIs_2}, one may follow the approach outlined in Ref.~\cite{Mutter2022} to obtain noise characteristics from the ANI in the linearly coupled case. Hence, the power spectral density can also be obtained in quadratically coupled qubit-resonator systems. Noise features of common noise types can be obtained by reducing $\llangle \mathcal{I}_2 \rrangle$ to $\llangle \mathcal{I}_1 \rrangle$ as well, even though it may often be easier to directly use the analytical expression for $\llangle \mathcal{I}_2 \rrangle$ given in Eqs.~\eqref{eq:ANI_2_white_noise} and~\eqref{eq:ANI_2_qs_noise} in practice.

\section{Conclusion}\label{sec:conclusion}
In this paper we extended the theory of noise spectroscopy based on the transient transmission to include quadratic qubit-resonator interactions, which occur in nanonmechanical and superconducting resonators coupled to solid-state qubits. Allowing for arbitrary initial qubit coherences, we derived and solved the quantum Langevin equations in the presence of a noisy qubit in first-order perturbation theory and showed that noise features are imprinted in the averaged resonator transmission and phase. Measuring both of these quantities permits the extraction of the ANI $\llangle \mathcal{I}_2 \rrangle$ for quadratically coupled systems. By converting $\llangle \mathcal{I}_2 \rrangle$ to the ANI for linearly coupled systems, one may extract the noise power spectral density from transmission measurements using established methods. Our results expand the scope of applications of transmission-based noise spectroscopy to a wide variety of systems including superconducting, charge and spin qubits coupled to electromagnetic cavities and nanomechanical resonators.

\section{Acknowledgments}
This research is supported by the German Research Foundation [Deutsche Forschungsgemeinschaft (DFG)] under Project No.~450396347 and No.~425217212 - SFB 1432.

\appendix
\onecolumngrid
\section{Real and imaginary parts of the transmission amplitude}
\label{appx:real_and_imaginary_parts_of_A}
In this Appendix we display the real and imaginary parts of the transmission amplitude in Eq.~\eqref{eq:transmission_amplitude_n} for both linear ($n=1$) and quadratic ($n=2$) qubit-resonator couplings. They read
	\begin{align}
	\begin{split}
		& \text{Re}(A_n)  =   - \frac{\sqrt{\kappa_1 \kappa_2 }}{\Delta_r^2  + \kappa^2/4} \bigg[ \frac{\kappa}{2} - e^{-\kappa t/2} \left( \cos (\Delta_r t) \frac{\kappa}{2} - \sin (\Delta_r t) \Delta_r  \right) + g n \sqrt{\Delta_r^2 + \frac{\kappa^2}{4}} \left( \frac{\sqrt{\kappa_1} \langle b_{\text{in}} \rangle }{\sqrt{\Delta_r^2  + \kappa^2/4}} \right)^{n-2} e^{-\kappa t/2} \\
		& \quad  \times \left\lbrace \left[ \text{Re} \langle \sigma_-^0 \rangle \text{Im}  \mathcal{I}_n  +  \text{Im} \langle \sigma_-^0 \rangle \text{Re}   \mathcal{I}_n  \right] \cos \left( [n-1] \zeta - \Delta_r t \right) + \left[ \text{Re} \langle \sigma_-^0 \rangle \text{Re}  \mathcal{I}_n  - \text{Im} \langle \sigma_-^0 \rangle \text{Im}   \mathcal{I}_n  \right]    \sin \left( [n-1] \zeta - \Delta_r t \right)  \right\rbrace  
		\bigg],
	\end{split}
	\end{align}
\begin{align}
	\begin{split}
			& \text{Im}(A_n)  =   - \frac{\sqrt{\kappa_1 \kappa_2 }}{\Delta_r^2  + \kappa^2/4} \bigg[ - \Delta_r +  e^{-\kappa t/2} \left( \sin (\Delta_r t) \frac{\kappa}{2} + \cos (\Delta_r t) \Delta_r  \right) + g n \sqrt{\Delta_r^2 + \frac{\kappa^2}{4}} \left( \frac{\sqrt{\kappa_1} \langle b_{\text{in}} \rangle }{\sqrt{\Delta_r^2  + \kappa^2/4}} \right)^{n-2} e^{-\kappa t/2} \\
		& \quad \times \left\lbrace \left[ \text{Re} \langle \sigma_-^0 \rangle \text{Im}  \mathcal{I}_n  +  \text{Im} \langle \sigma_-^0 \rangle \text{Re}   \mathcal{I}_n  \right] \sin \left( [n-1] \zeta - \Delta_r t \right) - \left[ \text{Re} \langle \sigma_-^0 \rangle \text{Re}  \mathcal{I}_n  - \text{Im} \langle \sigma_-^0 \rangle \text{Im}   \mathcal{I}_n  \right]    \cos \left( [n-1] \zeta - \Delta_r t \right)  \right\rbrace  
		\bigg],
	\end{split}
	\end{align}
where as in the main text we use the notation $\zeta = \arctan (2 \Delta_r/\kappa)$.
\twocolumngrid

\bibliography{charge_noise_cQED}
\end{document}